\documentstyle[preprint,aps]{revtex}
\begin{document}
\draft
\title{
Electronic structure and lattice relaxation
related to Fe in MgO
}
\author{ M.~A.~Korotin\cite{*}, A.~V.~Postnikov\cite{*},
         T.~Neumann, and G.~Borstel}
\address{
Universit\"at Osnabr\"uck -- Fachbereich Physik,
D-49069 Osnabr\"uck, Germany
}
\author{ V.~I.~Anisimov}
\address{
Institute of Metal Physics,
Russian Academy of Sciences, Yekaterinburg GSP-170, Russia
}
\author{M.~Methfessel}
\address{
Institut f\"ur Halbleiterphysik, P.O.Box 409,
D-15204 Frankfurt an der Oder, Germany
}
\date{\today}
\maketitle
\begin{abstract}
The electronic structure of Fe impurity in MgO was calculated
by the linear muffin-tin orbital--full-potential method
within the conventional local-density approximation
(LDA) and making use of the LDA+$U$ formalism. The importance
of introducing different potentials, depending on the screened
Coulomb integral $U$, is emphasized for obtaining a physically
reasonable ground state of the Fe$^{2+}$ ion configuration.
The symmetry
lowering of the ion electrostatic field leads to the observed
Jahn--Teller effect; related ligand relaxation confined
to tetragonal symmetry has been optimized based on the
full-potential total energy results. The electronic structure
of the Fe$^{3+}$ ion is also calculated and compared with
that of Fe$^{2+}$.
\end{abstract}


\narrowtext
\section{INTRODUCTION}
\label{sec:intro}

The electronic structure of defects in MgO has been the subject of
a number of theoretical studies.
Among the latest first-principles
calculations, one should mention the study on
anion vacancies by the muffin-tin Green's function
method\cite{klein87} and in a mixed-basis pseudopotential
supercell calculation\cite{wang90}.
In Ref.\cite{timmer}, the electronic structure
of Fe, Co, and Ni impurities
in MgO has been calculated making use of the Green's-function method
within the linearized-muffin-tin-orbitals (LMTO) --
atomic-spheres-approximation (ASA) formalism. In this paper,
the energy positions of the impurity $3d$ levels
in the gap have been calculated for several charge states
of the impurity ions. However, the effects of lattice relaxations
around cation impurity atoms have been ignored (although the
uniform relaxation around vacancies and substitutional hydrogen
has been considered
in Ref.\cite{wang90}), and the main emphasis has been on
the peculiarities of the electronic structure in these
defect systems.

The present study is motivated by the fact that Fe impurity
(as well as many other 3$d$ ions) in MgO
is a Jahn--Teller system, as has been observed
in electron spin resonance experiments (see, e.g.,
Refs.\cite{ham} and \cite{bates}).
However, the microscopic pattern of the underlying Jahn--Teller
distortion is not clear from the experiment, nor did the
{\it{ab initio}} calculations performed up to now succeed in
incorporating the essential effects of the interaction between
orbital moment of the impurity ion and the displacements
of ligand atoms. Based on an {\it{ab initio}} electronic
structure calculation scheme with no shape approximation
for the potential, we attempted to study the effect of
symmetry-lowering atomic displacements on the total energy
and thus to optimize the positions of ligand atoms towards
their relaxed local configuration around the impurity.
In doing this, we realized the importance of going beyond
the local density approximation (LDA)
in describing the electronic states
at the impurity site.

In Sec. \ref{sec:method}, we briefly specify
the calculation scheme we used.
In Sec. \ref{sec:LDA}, the results of a conventional
calculation for the Fe:MgO system within the
local-density approximation are outlined.
In Sec. \ref{sec:LDA+U}, we describe the effect
of introducing the orbital-dependent potential
within the so-called LDA+U approximation.
In Sec. \ref{sec:geomet}, the lattice relaxation
around Fe in MgO is analyzed, basing on total-energy
calculations for various geometries.
In Sec. \ref{sec:fe3+}, we present a simplified way
to model a metastable Fe$^{3+}$ ion configuration
of the impurity in MgO and analyze the electronic structure
corresponding to this configuration.

\section{CALCULATION METHOD}
\label{sec:method}

We performed the calculations for the 2$\times$2$\times$2
supercell of MgO (16 atoms), with one Mg atom substituted
by Fe, making use of the full-potential LMTO method\cite{msm1,msm2}.
The space is divided into nonoverlapping muffin-tin spheres,
the radii of which were chosen to be 1.65 a.u. for Mg (or Fe)
and 2.13 a.u. for O.
Mg~2$p$ and Fe~3$p$ states were treated as semicore,
and all higher-lying states included in the valence band panel.
Following the usual routine of
geometry-optimizing full-potential calculations, we calculated
the equilibrium lattice constant of the undistorted NaCl-type
crystal lattice of perfect MgO, which was found to be
98\% of the experimental value 4.21 \AA.
The band structure calculated for perfect MgO was found
to be in agreement with earlier
calculations\cite{klein87,wang90,timmer}. The valence band
is formed mainly by O~2$p$ states and the conduction band
by Mg~3$p$ states, with the gap separating them equal to
0.43 Ry (5.85 eV). This value is somewhat bigger than
in some other calculations, e.g. 4.36 eV\cite{wang90},
4.65 eV\cite{klein87}, 4.98 eV\cite{timmer}, but well below
the experimental estimate of about
7.8 eV\cite{roessler,williams,rao,fiermans}, which is a typical
error in calculating the band gap value within the LDA.

The Brillouin zone integration in the course of iterations
was performed over a mesh of special points;
total-energy convergency over the {\bf k} points has been
achieved at 10 special points for the main valence
band panel in the irreducible part of the Brillouin zone
for tetragonal symmetry (most of the calculations have been
performed in tetragonal symmetry for the reasons discussed
below). Densities of states were calculated
by the tetrahedron method with 75 {\bf k} points
in the irreducible part.

\section{IRON IMPURITY IN MgO: ELECTRONIC STRUCTURE IN THE LDA}
\label{sec:LDA}

Fe impurities which substitute 1/8 of Mg atoms in the supercell
give rise to narrow levels in the gap above the mostly O~2$p$-type
valence band. In conventional LDA calculations, the potential
acting on all $d$ electrons of the same spin
at Fe site is identical, and
splitting of the impurity levels may occur due to spin polarization
or the crystal field effect.

The interaction between nearest Fe atoms, which are
separated only by one Mg atom in the [110] direction, gives rise to
the formation of narrow $t_{2g}$ and $e_g$ bands split by the crystal
field by $\sim$0.1 Ry.
The lower-lying $t_{2g}$ states
become fully occupied by 3 spin-up and 3 spin-down electrons,
and the Fe impurity is stripped of its 4$s$ electrons,
thus being left in the $^1A_1$ atomic configuration
with ionicity 2+. The Fermi level
falls, therefore, in the gap between the $t_{2g}$ and $e_g$ bands.

The density of states obtained by the full-potential LMTO
calculation, in the option which incorporated spin-resolved
potential and charge density within the Fe MT sphere,
but in fact resulted in a zero spin moment,
is shown in Fig.\ \ref{dos_lda}.
A nonmagnetic configuration occurs quite seldom for
Fe impurities, and in the present case the calculated result
clearly contradicts the experimental observation that the
ground state configuration of Fe in MgO is $^5T_2$
(2$e_g$, 4$t_{2g}$) -- see the discussion in Ref.\cite{timmer}.
Therefore we were careful in additionally investigating
the question of whether the obtained non-magnetic
configuration of Fe is a drawback of LDA, or an artifact
due to some particular cell geometry, choice of sphere
radii, etc. In a series of calculations by the tight-binding (TB) LMTO
method\cite{tblmto} it was proven that the first assumption is correct,
since convergence to the same nonmagnetic configuration
occured irrespective of the cell size (as big as a 32-atom
supercell of MgO:Fe has been tested), presence of empty
spheres, and other starting conditions.
The (0$e_g$, 6$t_{2g}$) configuration turned out to be very stable,
as any attempt to remove electrons from $t_{2g}$ states to
$e_g$ states energetically quite separated from them
led to the situation when the Fermi level crosses two narrow
peaks (spin-down $t_{2g}$ and spin-up $e_g$) and is therefore
energetically very unfavorable.

The converged
density of states determined for the 32-atom supercell
by the TB-LMTO method was almost identical to that shown
in Fig.\ \ref{dos_lda}, with the same position of $t_{2g}$
and $e_g$ impurity peaks in the energy gap, only that
the widths of these peaks have been
substantially reduced.
We considered this similarity as an indication that
our results from a full-potential calculation in
regard to the structure optimization may be
probably only slightly corrected by further increase
of the supercell size.

An obvious drawback of conventional LDA calculations
performed for potentially magnetic impurities in insulators
is the implicit orbital degeneracy within the $d$ shell.
In reality, one would expect that the potential acting on
any particular electronic state depends on whether this state is
occupied or empty. Being effectively of just marginal effect
for strongly hybridized systems, this discrepancy becomes
quite important as the degree of localization increases.
This is exactly the case in the MgO:Fe system, and we discuss
below how the results could be amended by lifting the
orbital degeneracy.

\section{RESULTS OF LDA+$U$ CALCULATIONS}
\label{sec:LDA+U}

In a simple way, the orbital- (and occupation-) dependent
potential is introduced in the so-called LDA+$U$
approach\cite{oxides2,nio93}. The correction to the potential
enters as a constant term, determined primarily by
the screened effective Coulomb integral $U$ and the screened
exchange parameter $J$ (see Ref.\cite{nio93}).

The first parameter for the system in question
has been estimated from self-consistent full-potential
calculation within the LDA
using the scheme of Gunnarsson {\em et al.}\cite{gun}
to be $U$=0.489 Ry.
The value of $J$ for Fe in MgO
is expected to be the same as in FeO
and was taken from Anisimov {\em et al.}\cite{oxides2}.

In an earlier application\cite{oxides2},
the LDA+$U$ method
has been applied to a number of oxide systems, making it possible
to obtain reasonable {\it{ab initio}} estimations
of the band gap value and magnetic moments that were
otherwise not possible within the LDA.
When applying this scheme to MgO, however, it was not possible
to correct the underestimated value of the band gap
in a similar way, because of rather weak localization
of the Mg~3$p$ states forming the conduction band. Therefore,
the effect of LDA+$U$ was confined in our case only to the
Fe impurity, for which the conventional LDA approach
predicts an essentially incorrect charge configuration.

On introducing the occupation-dependent potential
acting on each Fe~3$d$ orbital independently as is
described above, the energy positions of the impurity
levels in the gap were drastically changed
(Fig.\ \ref{dos_lda+u}). As the filled and empty states
of $t_{2g}$ symmetry are no longer forced to reside
at the same energy position, as was the case in the LDA
calculation, the system finds its way in the course
of iterations to a more stable configuration where
the Fermi level falls in the gap between filled
($xy$) and empty ($xz$,$yz$) states of $t_{2g}$
symmetry and thus the Hund's rule is satisfied.
Majority-spin $e_g$ states are then completely filled,
in contrast to what was predicted by the LDA calculation,
and the atomic configuration becomes
(2$e_g$, 4$t_{2g}$).

It was essential to perform the Brillouin zone integration
over at least 1/16, i.e. over the irreducible wedge
corresponding to the tetragonal structure, and not over
1/48 as for cubic symmetry, in order to account for
the difference which appears now between $xy$ and
($xz$, $yz$) states which are differently occupied.
This does not mean that we force the system to split the $d$
state in this particular way. The use of the
appropriate symmetry is just a means to save effort
in the integration over the whole Brillouin zone, which
would give the same scheme of splitting for the $d$ states.

It is seen from Fig.\ \ref{dos_lda+u} that
the degeneracy pattern
of Fe~3$d$ levels in the ligand field of nominally
cubic symmetry looks like that corresponding to a
tetragonal crystal field, because of the symmetry-lowering
effect of filled $xy$ states.
As is known,
$t_{2g}$ states should split into $b_{2g}$ and $e_g$,
and this is consistent with the remaining degeneracy
of $xz$ and $yz$ states. Then, $e_g$ should split
into $a_{1g}$ and $b_{1g}$, which is seen as the lifted
degeneracy between (3$z^2-1$) and ($x^2-y^2$) states;
however, they both are either filled (for majority
spin) or empty (for minority spin) and therefore are acted on by
the same $U$-dependent correction to the potential.

The tetragonal-symmetry electrostatic field at the Fe site
may affect the relaxation of the ligand atoms around the
impurity. In addition to the totally symmetric
uniform relaxation which
was studied, e.g., around vacancies in Ref.\cite{wang90},
we should now consider the effect of symmetry-lowering tetragonal
displacements of oxygen atoms. The results of corresponding
total-energy calculations are discussed below.

\section{OPTIMIZATION OF THE LIGAND GEOMETRY}
\label{sec:geomet}

Not much is known about the structural microscopic configuration
of atoms in the MgO crystal around the Fe impurity, apart
from the mere fact that this system exhibits dynamic Jahn--Teller
behavior. The origin of the Jahn--Teller displacement
seems to be essentially the tetragonal electrostatic field
related to the ground-state configuration of the Fe$^{2+}$
ion as discussed above, and the dynamical character
reveals merely an exchange between three equivalent
directions of the tetragonal axis in the initially cubic crystal.

It may be expected that the strongest effect
on the total energy will be from the totally symmetric uniform
adjustment of ligand shells as the Mg$^{2+}$ ion is
substituted by Fe$^{2+}$.
On top of this uniform relaxation, the Jahn--Teller effect
is expected to emerge in the form of symmetry lowering due to
additional displacements of ligand atoms consistent with
the tetragonal symmetry.
In order to reduce the computational effort
for examining the effect of various independent atomic
displacements on the total energy, we restricted ourselves
to considering the displacement of the nearest oxygen atoms only.
That means that we considered only uniform breathing
of the O$_6$ octahedron and the additional axial displacement
of its two apical atoms.

The total energy as a function of uniform O$_6$ breathing
is shown in Fig.\ \ref{etot}. Open circles mark the
energy values obtained in the LDA+$U$ calculation,
and closed triangles the results from the conventional
LDA calculation, with nonmagnetic configuration of the
Fe$^{2+}$ ion. The latter values correspond to other
absolute values of the total energy, but the plot is
shifted so as to coincide with the LDA+$U$ curve for
zero breathing, for better comparison. One can see that
the outward expansion of the octahedron by $\sim$1\%
of the lattice constant
(1.15\% in the LDA+$U$ calculation) is obtained in the
calculation in both cases, so it is mostly a purely
electrostatic effect, related to the different electronegativity
of Fe and Mg. However, the magnitude of the energy lowering
is substantially larger
in the LDA+$U$ calculation.

Subsequent squeezing of the O$_6$ octahedron
along [001] by 0.5\% of the lattice constant lead
to further decrease of the total energy.
The corresponding densities of states (DOS's)
obtained in the LDA+$U$ calculation
are presented in
Fig.\ \ref{dos_lda+u}.

\section{CALCULATION FOR Fe$^{3+}$ ION}
\label{sec:fe3+}

Although substitutional Fe in MgO exists mostly as a 2+ ion,
the Fe$^{3+}$ configuration is also known to exist, at least
as a metastable state. It was pointed out
by Schultheiss\cite{fe3+} that upon irradiation with x rays or
electrons, Fe$^{3+}$ ions are formed
due to the trapping of holes at Fe$^{2+}$ sites.
Since the origin of the holes
to be trapped is most probably related to the lack of Mg atoms,
an adequate modeling of such a defect should in principle
incorporate this effect somehow in the construction of the
supercell. In a simplified attempt to get some insight
into the rearrangement of electronic states at the Fe site
due to the absorption of an extra hole, we performed a
self-consistent LDA+$U$ calculation with the option that the Fermi level
was searched for at one electron less per supercell
than should exist from the electroneutrality condition.
This means physically that an electron is removed from the
Fe site (since the band just below the Fermi level
is of essentially Fe~3$d$ character), and no charge compensation
is nominally introduced.
If we consider then the possible effect of restoring
the charge neutrality in the simplest form, i.e., due
to the introduction of compensatory charge uniformly
distributed over the crystal, it will obviously lead
merely to a slight rigid shift of all bands which
have been calculated as described above.
Therefore this simplified approach
will be probably of no use for the study of relaxation where
the spatial distribution of the compensating charge is
really important, but it is expected to provide an adequate
description of electronic states at the impurity site.

\begin{figure}
\caption{
Total DOS per unit cell
and local orbital-resolved DOS on the Fe site
for the Fe$^{3+}$ configuration in
FeMg$_7$O$_8$.
}
\label{dos_3+}
\end{figure}
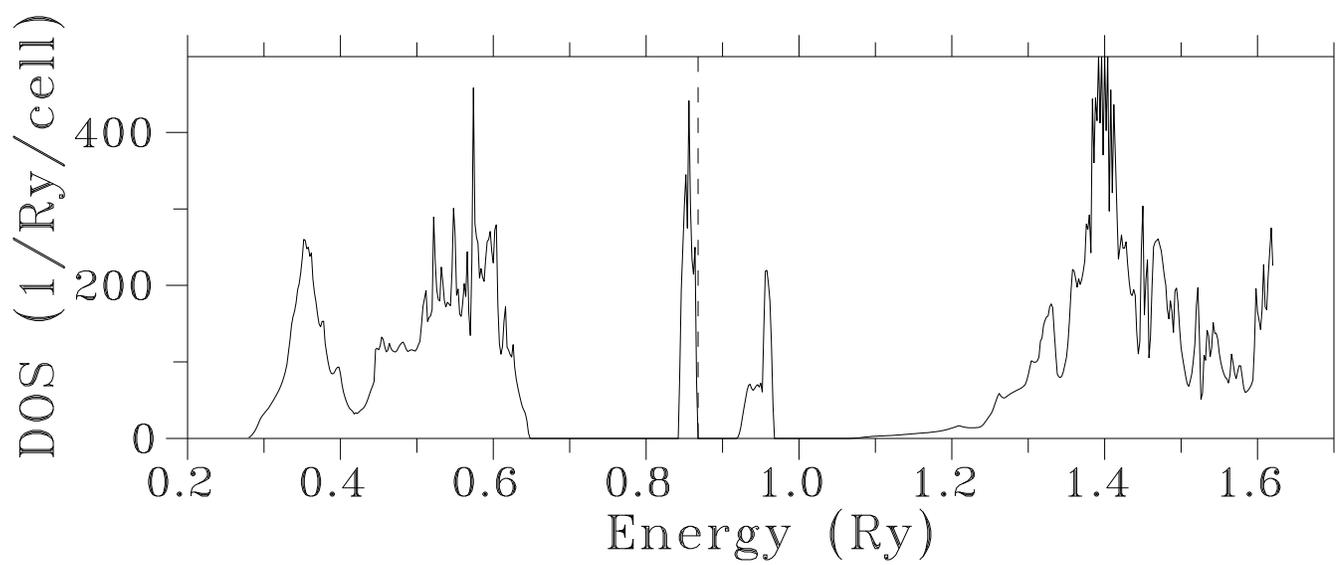
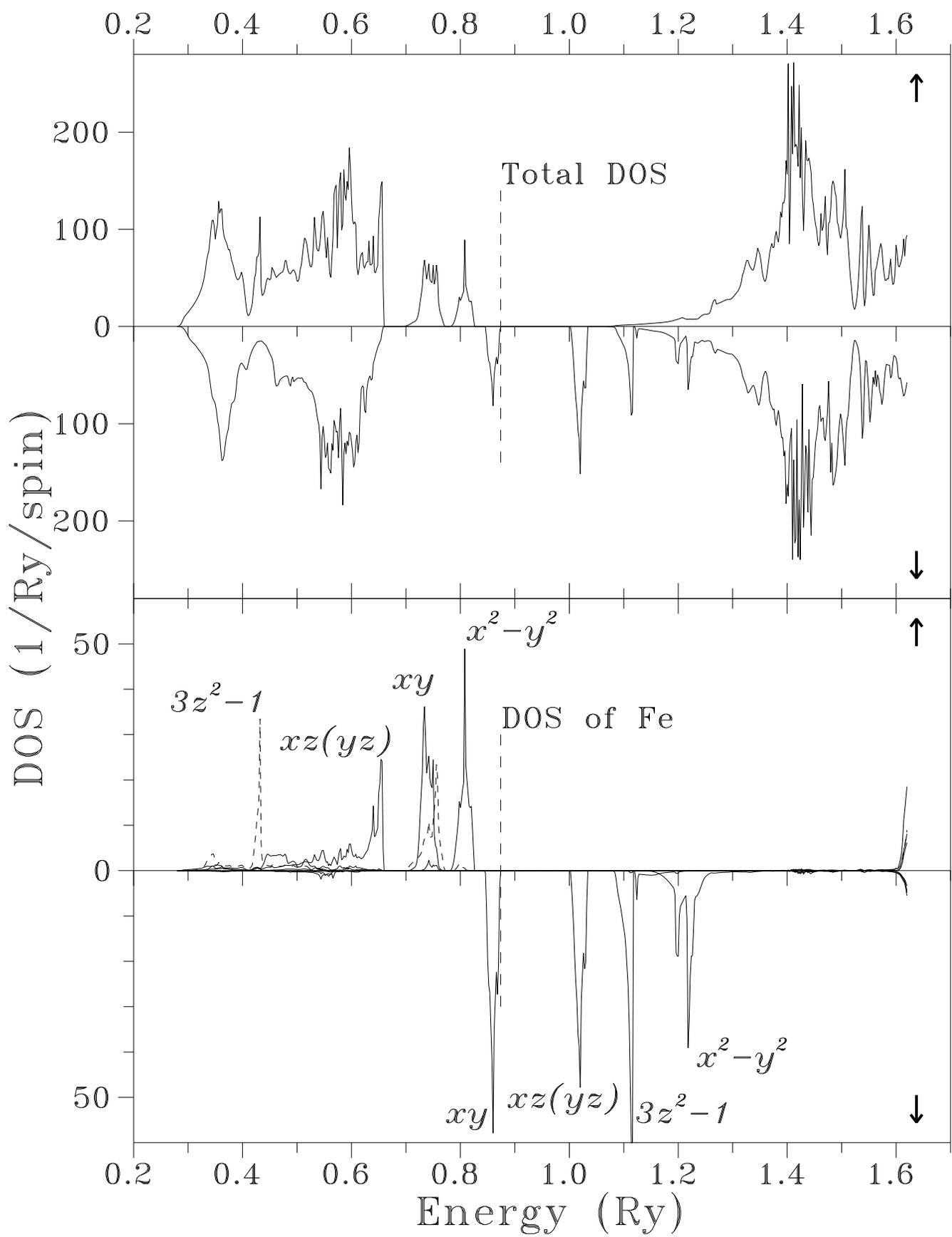
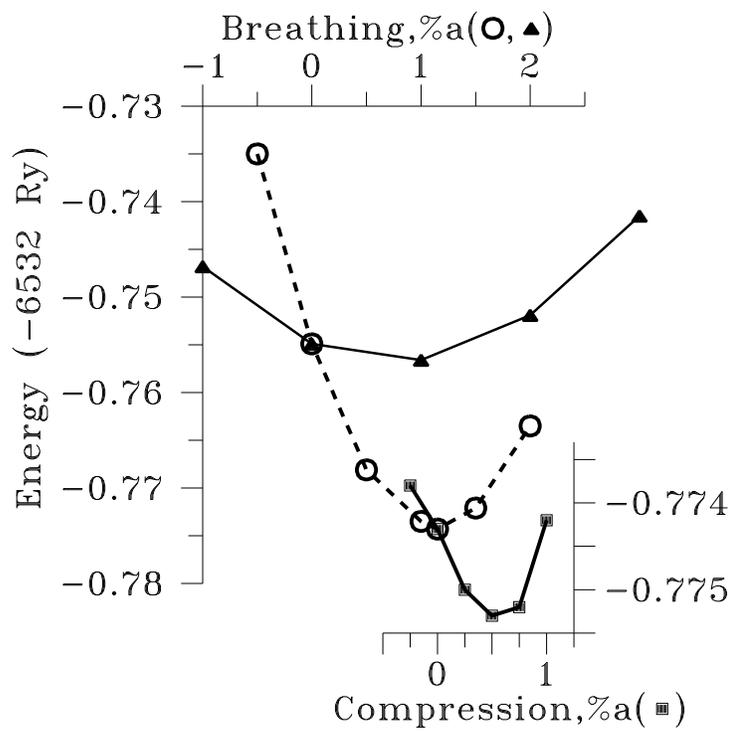
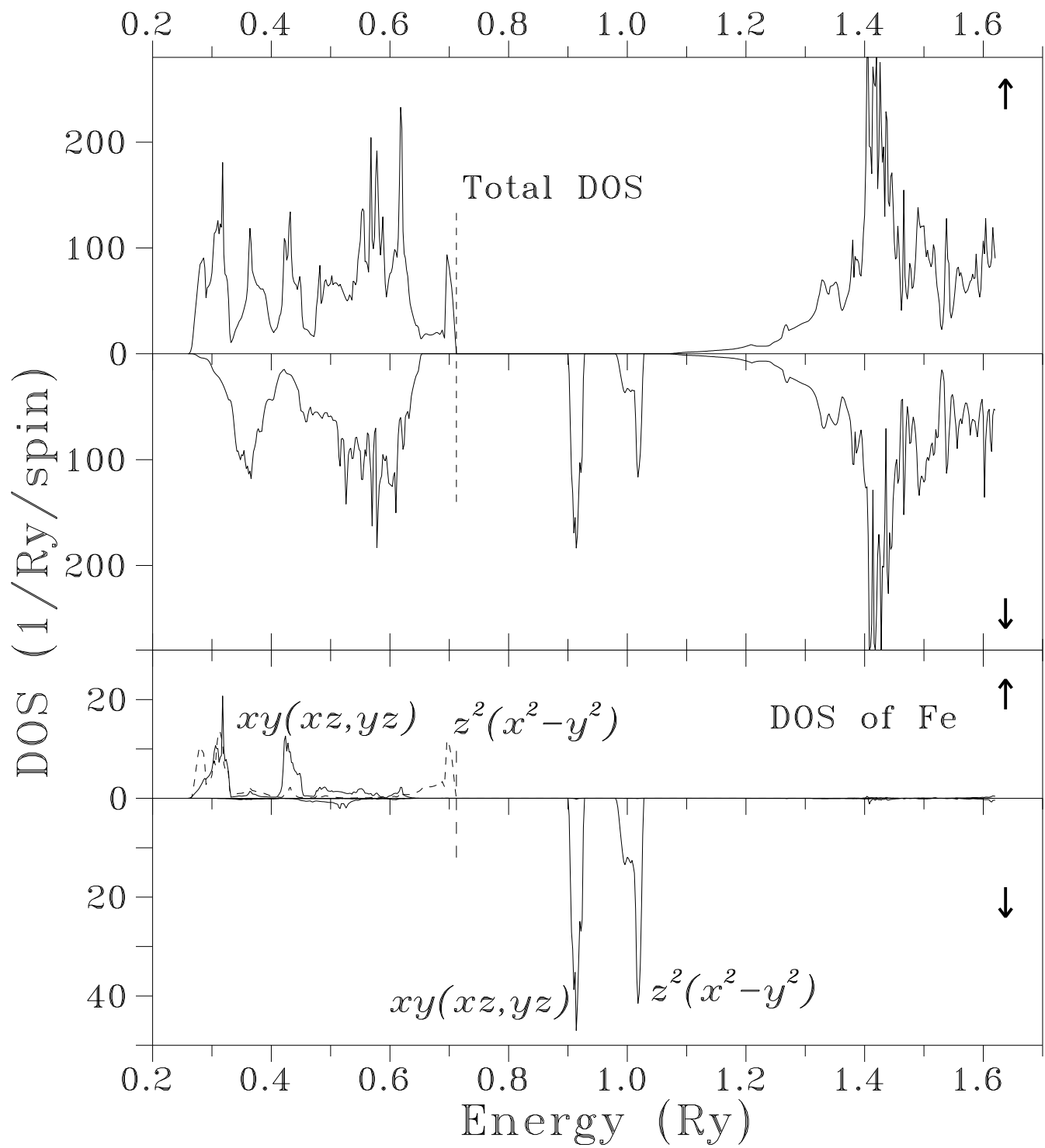

The resulting DOS is shown in Fig.\ \ref{dos_3+}.
It is rather different from the DOS of the Fe$^{2+}$ ion,
and the origin of this difference is that as the minority-spin
$xy$ state, i.e., the highest occupied one, is stripped of
its electron, it becomes equivalent, with respect to the
potential acting on them, to $xz$ and $yz$ states,
and the threefold degeneracy of the $t_{2g}$ state is
restored in the course of iterations, even as we solved
the band structure problem in nominally tetragonal symmetry.
At the same time (as the half-filled $d$ shell possesses full
cubic symmetry), the twofold degeneracy of $z^2$ and $x^2-y^2$
states is restored. Larger exchange splitting,
corresponding now to almost 5$\mu_B$, shifts Fe majority-spin
states down within the valence band region, and crystal-field-split
minority-spin states are situated now
at $\sim$0.2 Ry and $\sim$0.3 Ry above the Fermi level.

\section{CONCLUSION}
\label{sec:conclu}

In the course of studying the electronic structure of Fe
in MgO from first principles,
we have found that an electronic-structure calculation
within the conventional LDA erroneously predicts
the ground-state configuration of Fe in MgO to be
nonmagnetic, with the (0$e_g$, 6$t_{2g}$) configuration.
In order to amend this drawback and to go beyond the LDA,
an occupation-dependent correction to the potentials
of different impurity states has been implemented
in the LDA+$U$ scheme. The resulting electronic
configuration of the Fe$^{2+}$ ion was found to be
consistent with the Hund's rule, and the lifting of
the orbital degeneracy between occupied and empty $d$ states
resulted in a local field of tetragonal symmetry
at the impurity site. This symmetry-lowering
affects the relaxation of the ligand atom around the impurity,
making it a combination of a 1.15\% (in units of the lattice
constant) totally symmetric outward expansion of the O$_6$
octahedron and a 0.5\% inward squeezing of two apical
oxygen atoms. For the Fe$^{3+}$ ionic configuration of
the impurity, the orbital splitting only persists between
$t_{2g}$ and $e_g$ states of the $d^5$ shell, and the local
cubic symmetry is retained.

\acknowledgements

Financial support of the Deutsche Forschungsgemeinschaft
is gratefully acknowledged.
AVP thanks G. Sawatzky
for useful discussions.
MAK and VIA gratefully acknowledge financial support
from the Netherlands NWO special fund for scientists from
the former Soviet Union.

\begin{figure}
\caption{
Total DOS per unit cell from the
FeMg$_7$O$_8$ supercell calculation (LDA result,
converged to nonmagnetic configuration).
The Fermi level is indicated by the vertical dashed line.
}
\label{dos_lda}
\end{figure}

\begin{figure}
\caption{
Spin-resolved total DOS per unit cell
and local orbital-resolved DOS at the Fe site
in FeMg$_7$O$_8$ for the optimized geometry (LDA+$U$ result).
}
\label{dos_lda+u}
\end{figure}

\begin{figure}
\caption{
Total energy (per 16-atom unit cell)
versus uniform breathing (circles)
and tetragonal squeezing (squares)
of the O$_6$ octahedron.
Triangles: effect of breathing on the total energy
in the LDA calculation (see text).
}
\label{etot}
\end{figure}

\end{document}